\documentclass[twocolumn,aps,prd,amsmath,amssymb]{revtex4}
\bibpunct{}{}{,}{s}{}{,} 

\begin{document}

\title{Diffeomorphism invariance in spherically symmetric loop quantum gravity}

\author{ Rodolfo Gambini$^{1}$,
Jorge Pullin$^{2}$}
\affiliation {
1. Instituto de F\'{\i}sica, Facultad de Ciencias, 
Igu\'a 4225, esq. Mataojo, Montevideo, Uruguay. \\
2. Department of Physics and Astronomy, Louisiana State University,
Baton Rouge, LA 70803-4001}

\begin{abstract}
We study the issue of the recovery of diffeomorphism invariance in the
recently introduced loop quantum gravity treatment of the exterior
Schwarzschild space-time. Although the loop quantization agrees with
the quantization in terms of metric variables in identifying the
physical Hilbert space, we show that diffeomorphism invariance in
space-time is recovered with certain limitations due to the use of
holonomic variables in the loop treatment of the model. This resembles
behaviors that are expected in the full theory.
\end{abstract}

\maketitle
\section{Introduction}

Interest in the application of loop quantum gravity technique to model
systems has significantly increased in the recent years.  Most
treatments have dealt with homogeneous space-times (see the 
recent reviews of Ashtekar and Bojowald\cite{lqc}). 
More recently extensions to the exterior\cite{exterior} and 
interior\cite{interior} Schwarzschild space-times
were carried out, and also with some limitations to Gowdy 
models\cite{Gowdy}.  In all these studies, as is customary in mini or
midi-superspace treatments, gauge fixings are conducted in order to
exploit the simplifications inherent in the symmetries present in the
models.  On the other hand, in the full theory, one is interested in
diffeomorphism invariance. It is therefore of interest to study in the
context of the symmetry reduced models what happens to any remnants of
diffeomorphism invariance that are left after the gauge fixing.  In
this paper we would like to discuss the issue of diffeomorphism
invariance within the context of the treatment of the exterior
spherically symmetric space-times.  We will show that the remaining
diffeomorphism invariance is successfully recovered in the
semi-classical limit of the quantum theory but that there are
limitations imposed by the used of holonomic variables in the
quantization. In spite of this, the quantum theory has the same
degrees of freedom as if one used metric variables and the 
solutions of the semiclassical ``polymerized'' theory are uniquely
determined by the mass, as in ordinary general relativity. This 
appears in contrast to treatments of the interior\cite{interior} 
of the Schwarzschild
space-time  and may shed light on the treatment of the
complete space-time where the issue of uniqueness is still not 
settled\cite{complete}.

The organization of this paper is as follows. In the next section
we briefly review the loop quantum gravity treatment of the exterior
of the Schwarzschild space-time. In section III we discuss the issue
of diffeomorphism invariance. We end with a discussion.

\section{Spherically symmetric space-times in loop quantum gravity}

We briefly review here the treatment of the exterior in loop quantum
gravity. More details can be found in our previous paper\cite{exterior}.

One assumes that the topology of the spatial manifold is of the form
$\Sigma=R^+\times S^2$. We will choose a radial coordinate $x$ and
study the theory in the range $[0,\infty]$. We will later assume that
there is a horizon at $x=0$, with appropriate boundary conditions as
we discuss below.
The invariant connection can be written 
as,
\begin{eqnarray}
  A &=& A_x(x) \Lambda_3 dx + 
\left(A_1(x) \Lambda_1+ A_2(x) \Lambda_2\right)d\theta\\&& +
\left(\left(A_1(x) \Lambda_2- A_2(x) \Lambda_1\right)\sin \theta +
\Lambda_3 \cos \theta\right) d\varphi,\nonumber
\end{eqnarray}
where $A_x, A_1$ and $A_2$ are real arbitrary functions on $R^+$,
the $\Lambda_I$ are generators of $su(2)$, for instance $\Lambda_I = 
-i\sigma_I/2$ where $\sigma_I$ are the Pauli matrices or
rigid rotations thereof. The invariant triad takes the form,
\begin{eqnarray}
  E &=& E^x(x) \Lambda_3 \sin \theta {\partial \over \partial x} + 
\left(E^1(x) \Lambda_1 + E^2(x) \Lambda_2\right) \sin \theta {\partial \over 
\partial \theta} \nonumber\\&&
+
\left(E^1(x) \Lambda_2 - E^2(x) \Lambda_1\right) {\partial \over 
\partial \varphi},
\end{eqnarray}
where again, $E^x, E^1$ and $E^2$ are functions on  $R^+$. 

As discussed in our recent paper\cite{exterior} and 
originally by Bojowald and Swiderski\cite{boswi}, 
it is best to make several changes of variables to simplify
things and improve asymptotic behaviors. It is also useful to gauge
fix the diffeomorphism constraint to simplify the model as much as
possible.  It would be too lengthy and not particularly useful to go
through all the steps here. It suffices to notice that one is left
with one pair of canonical variables $E^\varphi$ and ${A}_\varphi$ (in
our recent paper\cite{exterior} called $\bar{A}_\varphi$), and that they are
related to the traditional canonical variables in spherical symmetry
$ds^2=\Lambda^2 dx^2+R^2 d\Omega^2$ by $\Lambda=E^\varphi/(x+a)$ and
$P_\Lambda= -(x+a)A_\varphi/(2\gamma)$ where $\gamma$ is the Immirzi
parameter and $P_\Lambda$ is the momentum canonically conjugate to
$\Lambda$. The gauge fixing chosen is such that $R=(x+a)$ where $a$
is, at the moment, a dynamical variable function of $t$. We will also
choose $A_\varphi$ to be independent of $t$. The variable $x$ ranges
from zero to infinity. At zero we will impose isolated horizon
boundary conditions, i.e. $x=0$ will be the horizon, whereas
$x=\infty$ corresponds to $i^0$. An asymptotic analysis of terms at
spatial infinity shows that $a$ ends up being a constant related to
the mass of the space-time. Again we refer the reader to
our recent paper\cite{exterior} for details.

In terms of these
variables the Hamiltonian constraint reads,
\begin{eqnarray}\label{hamilclass}
  H&=&-{E^\varphi \over (x+a) \gamma^2}\left({A^2_\varphi (x+a)\over 8}\right)'
-{E^\varphi \over 2 (x+a)}\\&&+ {3 (x+a) \over 2 E^\varphi} 
+ (x+a)^2 \left({1 \over E^\varphi}\right)'=0.\nonumber
\end{eqnarray}
and since the variables are gauge invariant there is no Gauss law.
The Hamiltonian has a
non-trivial Poisson bracket with itself, proportional to a Hamiltonian with
structure functions. This makes the treatment of the constraint at a 
quantum level problematic since it has the usual ``problem of dynamics''
(see Giesel and Thiemann\cite{thiemanngiesel} for a good discussion) . 
To avoid this in a first approach, it is worthwhile noticing that through a
simple rescaling, the Hamiltonian constraint can be made Abelian,
just multiplying by $\frac{2(x+a)}{E^\varphi}$ and grouping terms as
\begin{equation}\label{abelianized}
H= \left(\frac{(x+a)^3}{(E^\varphi)^2}\right)'-1 -\frac{1}{4 \gamma^2}
\left((x+a) A_\varphi^2\right)'=0,
\end{equation}
yields and Abelian constraint. Since the constraint is a total
derivative, it can immediately be integrated to yield,
\begin{equation}
\int H dx = C =
\left(\frac{(x+a)^3}{(E^\varphi)^2}\right)-x -\frac{1}{4 \gamma^2}
\left((x+a) A_\varphi^2\right),
\end{equation}
with $C$ a constant of integration. Recalling that at $x=0$ the
isolated horizon boundary conditions imply $1/E^\varphi=0$ and
$A_\varphi=0$ one gets that the constant of integration $C$ vanishes.
This in particular implies that at infinity, $a=2M$, imposing the
appropriate boundary conditions there, $E^\varphi=x+3M$, 
$A_\varphi=0$.

To promote the constraint to a quantum operator, one needs to discretize
the radial direction and then apply techniques at each point akin to those
of loop quantum cosmology.
One wishes to write the discretization in terms of classical quantities
that are straightforward to represent in the quantum theory. Here one
has to make choices, since there are infinitely many ways of
discretizing a classical expression. In particular, we will notice
that there exists, for this model, a way of discretizing the
constraint in such a way that it remains first class (more precisely,
Abelian) upon discretization. This is unusual, and we do not expect
such a behavior in more general models.

We now proceed to discretize this expression and to ``polymerize'' it,
that is, to cast it in terms of quantities that are easily representable
by holonomies,
\begin{eqnarray}
  H^\rho_m &=& \frac{1}{\epsilon}\left[
\left(\frac{(x_m+2M)^3 \epsilon^2}{(E^\varphi_m)^2}-
\frac{(x_{m-1}+2M)^3\epsilon^2}{(E^\varphi_{m-1})^2}\right)
-\epsilon \right.\nonumber\\
&&-\frac{1}{4 \gamma^2 \rho^2}
\left(
(x_m+2M)\sin^2\left(\rho A_{\varphi,m}\right)\right.\nonumber\\
&&\left.\left.-(x_{m-1}+2M)\sin^2\left(\rho A_{\varphi,m-1}\right)
\right)
\right],
\end{eqnarray}
expression that recovers (\ref{abelianized}) in the limit $\epsilon\to
0$, $\rho \to 0$. In the above expression $x_m$ are the positions of
the lattice points and $\epsilon$ is the separation of two points in a
fiducial metric. Although it is not necessary, for simplicity we
assume $\epsilon$ is a constant. The parameter $\rho$ arises in the
``polymerization'', i.e. in replacing $A_{\varphi,m}$ by $\sin(\rho
A_{\varphi,j})/\rho$. Whereas the parameter $\epsilon$ is introduced
just as a calculational device and can be taken $\epsilon\to 0$ in the 
end, the parameter $\rho$ is expected in loop quantum gravity to have
a fundamental minimum value related to the quantum of area. 
The above expression is immediately Abelian since it
can be written as the difference of two terms, one dependent on the
variables at $m$ and the other at $m-1$. Therefore each term has
automatically vanishing Poisson brackets with itself and with the
other.

 To implement the constraints as quantum operators as one does in the
 Dirac procedure, it is convenient to solve the constraint for the
 $E^\varphi_m$,
\begin{equation}\label{convenient}
  E^\varphi_m = \pm \frac{(x_m+2M)\epsilon}{
\sqrt{1-\frac{2M}{x_m+2M}+\frac{1}{4\gamma^2\rho^2}
\sin^2\left( \rho  A_{\varphi,m}\right)}},
\end{equation}
and this relation can be immediately implemented as an operatorial
relation and find the states that satisfy it. It should be noted that
this relation can be implemented for other gauges as well in a
straightforward manner. The states are given by,
\begin{equation}\label{109}
  \Psi[A_{\varphi,m},\tau,M] =C(\tau,M) \exp\left(
\pm \frac{i}{\ell_{\rm Planck}^2}\sum_m
f[A_{\varphi,m}]\right), 
\end{equation}
where $C(\tau,M)$ is a function of the variables at the boundary
$\tau$ and $M$, which has to solve the constraint at the boundary, as
we shall soon see. $\tau$ is the proper time at infinity, that for
instance determines the position of the spatial hypersurfaces of
vanishing extrinsic curvature (usual Schwarzschild slicings).  The
functional $f$ has the form,
\begin{eqnarray}
&&  f[A_{\varphi,m}]=
\frac{1}{4\gamma^2\rho^2\left(1-\frac{2M}{x_m+2M}\right)}(x_m+2M)
\nonumber \epsilon\\&&
\left[F\left(\sin( \rho A_{\varphi,m}),
\frac{i}{4\gamma^2\rho^2\left(1-\frac{2M}{x_m+2M}\right)}\right)\right.\\
&& \left.+ 2 F\left(1,
\frac{i}{4\gamma^2\rho^2\left(1-\frac{2M}{x_m+2M}\right)}\right)
{\rm sgn}\left(\sin(\rho A_{\varphi,m})\right)\right] \nonumber
\end{eqnarray}
with $F(\phi,m)\equiv \int_0^\phi (1-m^2\sin^2 t)^{-1/2}dt$ the Jacobi
Elliptic function of the first kind. Notice that the continuum limit
of this expression for the state is immediate, i.e. the sum in $m$ 
becomes an integral.

We now need to impose the constraints on the boundary, in particular
$p^\tau = -M$ (in the limit $N\to\infty$). Quantum mechanically
$\hat{p}^\tau = -i \ell_{\rm Planck}^2 {\partial/\partial \tau}$ and
therefore,
\begin{equation}
  C(\tau, M) = C_0(M) \exp\left(-\frac{i M \tau}{\ell_{\rm Planck}^2}\right)
\end{equation}
and $C_0(M)$ is an arbitrary function. This is analogous to the
quantization that Kucha\v{r} found where one had wavefunctions that
only depended on the mass. We have therefore completely solved the
theory.

\section{Diffeomorphism invariance of the model}


We start by pointing out that the quantization is straightforward,
since the only remaining canonical variables are $M$ and $\tau$. These
variables have no dynamics. 
One can immediately introduce an eigenbasis of the mass operator,
labeled by eigenvalues $m$, 
$\hat{M} \phi(m)=m \phi(m)$ and the equations of motion at the boundary
imply that the $\phi(m)$ do not evolve. This completes the quantization.

Since we have isolated the true degree of freedom of the model and
quantized it, there are no remnants left of the diffeomorphism
invariance of space-time in any manifest way. To reconstruct
diffeomorphism invariance in an explicit form it is useful to
introduce evolving constants\cite{evolving}. For instance, given that
the mass of the space-time can be written as a function of the
canonical variables $M=M(E^\varphi,\hat{A}_\varphi)$, one can
construct an evolving constant associated with the triad as
$E^\varphi_{\rm Evolv}=E^\varphi(M,A_\varphi^{(0)})$ where
$\hat{A}_\varphi^{(0)}$ is a parameter, as given by equation
(\ref{convenient}). Explicitly,
\begin{equation}\label{polymerized}
E^\varphi_{\rm Evolv} = \pm \frac{(x+2M)}{
\sqrt{1-\frac{2M}{x+2M}+\frac{1}{4\gamma^2\rho^2}
\sin^2\left(\rho A_{\varphi}^{(0)}\right)}}.
\end{equation}
 The quantity is such that if
one chooses ${A}_\varphi^{(0)}={A}_\varphi$ one recovers the
dynamical variable $E^\varphi$. The evolving constant is a Dirac
observable of the theory and therefore can be realized as an operator
acting on the physical space of the theory. Notice that the choice 
$A_{\varphi}^{(0)}=0$ corresponds to the ordinary form of the Schwarzschild
metric in Schwarzschild coordinates.

The four dimensional metric of the model can be written in terms of 
$E^\varphi_{\rm Evolv}$ and the parameter $A_\varphi^{(0)}$,
by determining the lapse and shift using the gauge fixing condition and
setting to zero the time derivatives of the variables.
Therefore the components of the four dimensional metric can also be
viewed as evolving constants. The explicit expressions
are,
\begin{eqnarray}
g_{00}^{\rm Evolv}(M,A_\varphi^{(0)})
&=&-\frac{x^2}{(E^\varphi_{\rm Evolv})^2}
+\frac{\sin^2(\rho A_\varphi^{(0)})}{4\rho^2\gamma^2}\\
g_{0x}^{\rm Evolv}(M,A_\varphi^{(0)})
&=&\frac{E^\varphi_{\rm Evolv} \sin(\rho A_\varphi^{(0)})}{2\rho \gamma x}\\
g_{xx}^{\rm Evolv}(M,A_\varphi^{(0)})
&=&\frac{(E^\varphi_{\rm Evolv})^2}{x^2}.
\end{eqnarray}
It is worthwhile pointing out that all of the above expressions are
readily promoted to quantum operators acting on the physical Hilbert
space simply substituting $M$ by $\hat{M}$.

The above results hold in the quantum polymerized theory. It is worthwhile
comparing them with the results in classical general relativity. The
expressions for the components of the metric in
traditional general relativity are,
\begin{eqnarray}
g_{00}&=&-\frac{x^2}{(E^\varphi)^2}+\frac{A_\varphi^2}{4\gamma^2}\\
g_{0x}&=&\frac{E^\varphi A_\varphi}{2\gamma x}\\
g_{xx}&=&\frac{(E^\varphi)^2}{x^2}.
\end{eqnarray}

It is instructive to substitute the explicit expression of the triad,
\begin{equation}
E^\varphi = \pm \frac{(x+2M)}{
\sqrt{1-\frac{2M}{x+2M}+\frac{A^2_{\varphi}}{4\gamma^2}}}
\end{equation}
Different choices of gauge correspond to different choices of $A_\varphi$
and these translate themselves in different coordinate choices for the 
four-metric. 
The explicit form of the four dimensional metric therefore is,
\begin{eqnarray}
g_{00}&=& -1+\frac{2M}{x+2M}\label{21}\\
g_{0x}&=&\frac{A_\varphi}{2\gamma\sqrt{1-\frac{2M}{x+2M}+\frac{A_\varphi^2}{4\gamma^2}}}\\
g_{xx}&=& \frac{1}{1-\frac{2M}{x+2M}+\frac{A_\varphi^2}{4\gamma^2}}.
\label{23}
\end{eqnarray}

Since spatial diffeomorphisms have been gauge fixed the only
diffeomorphisms left are space-time ones, which modify the value of
$g_{xx}$ and $g_{0x}$.  If one starts in a gauge where
$A_\varphi=0$ with coordinates $t,x$, one can go to an arbitrary gauge
$A_\varphi(x)$ by choosing $x'=x$ and $t'=t-u(x)$ with
\begin{equation}
u(x)= \int_x^\infty dx 
\frac{A_\varphi(x)}
{2\gamma x \left(1-\frac{2M}{x+2M}+\frac{A_\varphi^2}{4\gamma^2}\right)}.
\label{change}
\end{equation}

Let us now compare with the quantum theory. In this example
things are so simple that we could actually talk about the full
quantum theory itself, it would just correspond to replace the mass by
a quantum operator in the following expressions.  Since the use of a
``polymerized'' classical theory to capture the semiclassical
behaviors of the quantum theory is a technique used in a variety of
contexts, we frame the discussion in it. The expression for $g_{00}$
is unchanged. The expressions for $g_{0x}$ and $g_{xx}$ become,
\begin{eqnarray}
g_{00}&=& -1+\frac{2M}{x+2M} \label{25}\\
g_{0x}&=&\frac{\sin(\rho A_\varphi)}{2\rho\gamma\sqrt{1-\frac{2M}{x+2M}+\frac{\sin\left(\rho A_\varphi\right)^2}{4\rho^2\gamma^2}}}\\
g_{xx}&=& \frac{1}{1-\frac{2M}{x+2M}+\frac{\sin\left(\rho A_\varphi\right)^2}{4\rho^2 \gamma^2}}.\label{27}
\end{eqnarray}
We therefore see that these expressions are particular cases of the
ones we found in the non-polymerized theory, in the sense that for every
choice of $A_\varphi$ for the polymerized theory one can find a choice in
classical general relativity that leads to the same metric. The converse,
however, is not true. The gauge transformations
of the polymerized theory therefore correspond to diffeomorphisms,
just like in classical general relativity. But not all of the
diffeomorphisms available in classical general relativity, at least
for finite values of $\rho$, appear in the polymerized theory.  It is
clear that if one chooses a small value of $\rho$ as suggested from
full loop quantum gravity, where it is associated with the quantum of
area which is related to Planck's length, ``most'' diffeomorphisms
will be allowed in the polymerized theory, but there will be a subset
that is not. The reason for this limitation is that one does not
expect diffeomorphism invariance to allow us to ``blow up'' regions
that are smaller than Planck size to macroscopic sizes. That would
imply that somehow we can probe space-time at sub-Planckian
lengths. This does not appear reasonable. We can see that this is
precisely what happens in this example. Which diffeomorphisms are
being excluded?  Comparing (\ref{21}-\ref{23}) to (\ref{25}-\ref{27})
we see that when $A_\varphi$ is large, this corresponds to $g_{xx}\to
0$ and $g_{0x}\to 1$ in classical general relativity, whereas in the
polymerized theory $g_{xx}$ reaches a minimum value. Let us recall
that we are talking about space-time diffeomorphisms here, that is,
changes of the space-time foliation (the radial coordinate is
fixed). That means one is excluding foliations where radial distances
become very small, i.e. foliations of large values of the extrinsic
curvature.

\section{Discussion}

The quantization of the exterior Schwarzschild space-time in loop
quantum gravity can be carried out completely and it isolates the same
true degrees of freedom as the quantization carried out by Kucha\v{r}
in terms of the traditional variables. One has that the only degree of
freedom is the mass of the space-time. Wavefunctions are functions of
the mass that do not evolve. In spite of this similarity, if one tries
to reconstruct space-time diffeomorphisms in terms of evolving
constants that are Dirac observables on the physical Hilbert space,
one notices effects due to the ``polymerization'' introduced by the
loop variables. In particular one notes that only a subset of
space-time diffeomorphisms get implemented. This corresponds physically
to the fact that one cannot probe distances of sub-Planckian nature.
This provides a simple example in a controlled situation 
of behaviors that are widely believed
to hold in the full theory.

It should be noted that in this paper we have taken the parameter
$\rho$ in the polymerized theory to be a constant. Current treatments
in cosmology suggest that an improved dynamics may be achieved with
$\rho$ that depends on the dynamical variables\cite{aspasi}. The
details of the conclusions about the permissible diffeomorphisms will
change if one makes such a choice, though we expect the generic
features to remain the same.

The issue of how diffeomorphisms get implemented in the polymerized
theory becomes quite relevant when one considers the full Kruskal-like
extension of the Schwarzschild space-time\cite{complete}. There, at
the moment, there exists knowledge of a family of solutions. It is not
clear if this family is unique or even if different members of the
family correspond to different space-times. This raises the issue of
the existence of a Birkhoff theorem in loop quantum gravity. In the
exterior case we have shown that the only quantum solutions can be
superpositions of space-times with different mass, in spite of the
fact that one does not implement in the semi-classical theory all the
diffeomorphisms present in the classical theory. In the interior case
it is known that the solutions of the ``polymerized'' semi-classical
theory may depend on an extra parameter in addition to the mass.  The
question is still open if an analysis of diffeomorphism symmetry like
we carried out in this paper can yield a Birkhoff-like theorem in the
case of the complete space-time.

\section{Acknowledgements}

We wish to thank Abhay Ashtekar for discussions and comments on the
manuscript.  This work was supported in part by grant NSF-PHY-0650715,
funds of the Hearne Institute for Theoretical Physics, FQXi, CCT-LSU
and Pedeciba (Uruguay).

\end{document}